\documentclass[reprint,pra]{revtex4-1}
\usepackage{amssymb,amsmath}
\usepackage{graphicx}
\usepackage{dcolumn}
\usepackage{multirow}
\usepackage{color}
\usepackage[english]{babel}

\begin{document}

\def\simlt{\mathrel{\lower .3ex \rlap{$\sim$}\raise .5ex \hbox{$<$}}}

\title{\textbf{\fontfamily{phv}\selectfont Disorder-induced valley-orbit hybrid states in Si quantum dots}}
\author{John~King~Gamble}
\author{M.~A.~Eriksson}
\author{S.~N.~Coppersmith}
\author{Mark~Friesen}
\affiliation{Department of Physics, University of Wisconsin-Madison, Madison, WI 53706}

\begin{abstract}
Quantum dots in silicon are 
promising candidates for implementation of solid-state quantum information processing.
It is important to understand the effects of the multiple conduction band valleys of silicon on the properties
of these devices.
Here we introduce a novel, systematic effective mass theory of valley-orbit coupling in disordered silicon systems.
This theory reveals valley-orbit 
hybridization effects that are detrimental for storing quantum information in the valley degree of freedom, including
non-vanishing dipole matrix elements
between valley states and altered intervalley tunneling.
\end{abstract}

\maketitle
Isolated electrons in semiconductor systems are a promising candidate for quantum computation because they exhibit excellent control and decoherence properties \cite{Hanson:2007p1217}.
Much recent progress has led to demonstrations of both spin- and charge-based qubits in GaAs 
\cite{Hayashi:2003p226804,Petta:2005p2180,Koppens:2006p766,Shulman:2012p202} and Si \cite{Morello:2010p687,Maune:2012p344,Shi:2012preprint}. While Si has better spin decoherence properties than GaAs \cite{Zwanenburg:2012preprint}, silicon's nontrivial conduction band valley structure is a complication 
\cite{Eriksson:2004p133}.

The presence of the valley degree of freedom in Si quantum dot devices
can lead to difficulty in isolating a two-state system to use as a qubit,
because
valley splitting energies can be the same order 
as both Zeeman splittings and orbital energy spacings \cite{Eriksson:2004p133,Goswami:2007p41,Culcer:2009p205302}. 
On the other hand,  it has been proposed to harness this valley degree of freedom to define noise-resistant qubits \cite{Eriksson:2004p133,Smelyanskiy:2005p081304,Culcer:2012p126804}.
Previous studies of valley states in Si have mainly focused on an idealized picture of the valley and orbital physics in which the system is taken to be disorder-free, and hence the valley and orbital degrees of freedom are good quantum numbers for the system \cite{Friesen:2007p115318,Baena:2012p035317}.

It has been recognized that structural disorder, 
such as atomic steps at the heterostructure interface, alloy disorder, or other types of correlated randomness, 
can introduce new effects such as intervalley tunneling \cite{Shiau:2007p195345,Culcer:2010p205315}.
Furthermore, recent experimental evidence for disorder-influenced valley-orbit physics has been found in both MOS \cite{Takashina:2006p236801,Yang:arXiv:1302.0983} 
and Si/SiGe systems \cite{shi:2011p233108,Shi:2012preprint}.
Studying disorder in silicon is especially challenging, since the conduction band valley states couple atomic-scale disorder to the micron-scale electron confinement that is typical of quantum dots.
To analyze this problem, researchers have used computationally intensive numerical techniques such atomistic tight binding \cite{Boykin:2004p165325,nielsen:2013p114304}, or analytical approaches that assume the effective mass theory holds with only minor corrections necessary \cite{Friesen:2006p202106,Friesen:2007p115318,Culcer:2010p205315}.

In this paper, we develop a systematic disorder-expansion technique that successfully reproduces the results of atomistic simulations, while retaining the appealing physical intuition and computational efficiency of effective mass theories.
Using this technique, in addition to the valley mixing matrix elements noticed previously, we identify matrix elements that correspond to valley-orbit hybridization, 
which were previously studied in an approximation using the two lowest energy $z-$states \cite{Friesen:2010p115324}.
We also show that the presence of these matrix elements leads to the emergence of effects not observed in 
previous analytical treatments. 
In particular, we show that disorder leads to finite dipole matrix elements between valley states,  
and quantitative corrections to intervalley tunneling.
Both effects are detrimental to quantum information stored in valley states.

\emph{Disorder-expansion effective mass theory.}---In Si quantum dots confined to a quantum well in the absence of disorder, one can show through tight-binding or effective mass theories that the energy
eigenstates 
$\Psi_{i,\pm}(\mathbf r) $ form
symmetric and antisymmetric valley doublets \cite{Friesen:2007p115318}:
\begin{equation}\label{unperturbed_solution}
\Psi_{i,\pm} (\mathbf r) = \frac{1}{\sqrt 2} \left[ u_{-k_0}(\mathbf r) e^{-i k_0 z} \pm u_{k_0}(\mathbf r) e^{i k_0 z} \right] h_i (\mathbf r),
\end{equation}
where $\mathbf r$ is the spatial position, $h_i$ is the electronic envelope function 
for the $i^{th}$ orbital, and
$u_{\pm k_0}$ is the periodic part of the Bloch function located at the conduction band minima $\mathbf k = \pm k_0 \hat z$,.
Here, $k_0 = 0.82 \cdot 2 \pi/a$ is the position of the valley minimum, and $a=0.543$~nm is the cubic lattice spacing in Si. 

To calculate the effects of disorder on valley states in Si accurately, researchers have previously
relied on atomistic tight-binding techniques \cite{Boykin:2004p165325,Kharche:2007p092109,Ahmed:2009preprint,nielsen:2013p114304}
which are numerical and extremely expensive computationally. Here, we introduce a new semi-analytical technique based on a systematic expansion in the matrix elements of disorder. This technique allows us to understand analytically and   compute accurately the effects of 
interface disorder
much faster than was previously possible. 

We consider an unperturbed problem consisting of a lateral, two-dimensional confinement potential $V(x,y)$
that describes a quantum dot or other device,
and a one-dimensional, vertical confinement potential $U(z)$ that includes the sharp interfaces, the quantum well barriers, and other slowly-varying components such as an applied electric field.
Since this problem is separable, the resulting wavefunction is written as $\Psi_{i,j} (\mathbf r) = F_i(x,y) \psi_j(z)$, where $i$ is the $x$-$y$ orbital index and
$j$ is the subband index \cite{Friesen:2007p115318}.

We solve the $x$-$y$ problem using the effective mass equation,
\begin{eqnarray}\label{xyHam}
\left[ -\frac{\hbar^2}{2 m_t} \left( \partial_x^2 +\partial_y^2\right) + V(x,y) \right] F_i(x,y) \nonumber\\
\equiv H^0_{xy}F_i(x,y) = \epsilon_i F_i(x,y) ~,
\end{eqnarray}
where 
$m_t = 0.19\, m_0$
is the transverse effective mass in Si, 
with $m_0$ the bare electron mass.
The $z$ problem is described similarly, with an unperturbed Hamiltonian $H_z^0$.
However, the solution method is different because $H_z^0$ includes sharp interface potentials that
couple different valley states.
There are a number of well-established techniques to solve this 1D problem,
including augmented effective mass treatments \cite{Friesen:2007p115318} and tight-binding techniques \cite{Boykin:2004p165325}, 
which yield solutions of the form noted in Eq.~(\ref{unperturbed_solution}). 
The results shown below are obtained using the 1D tight-binding treatment. 

\begin{figure}[tb!]
\includegraphics[width=8cm]{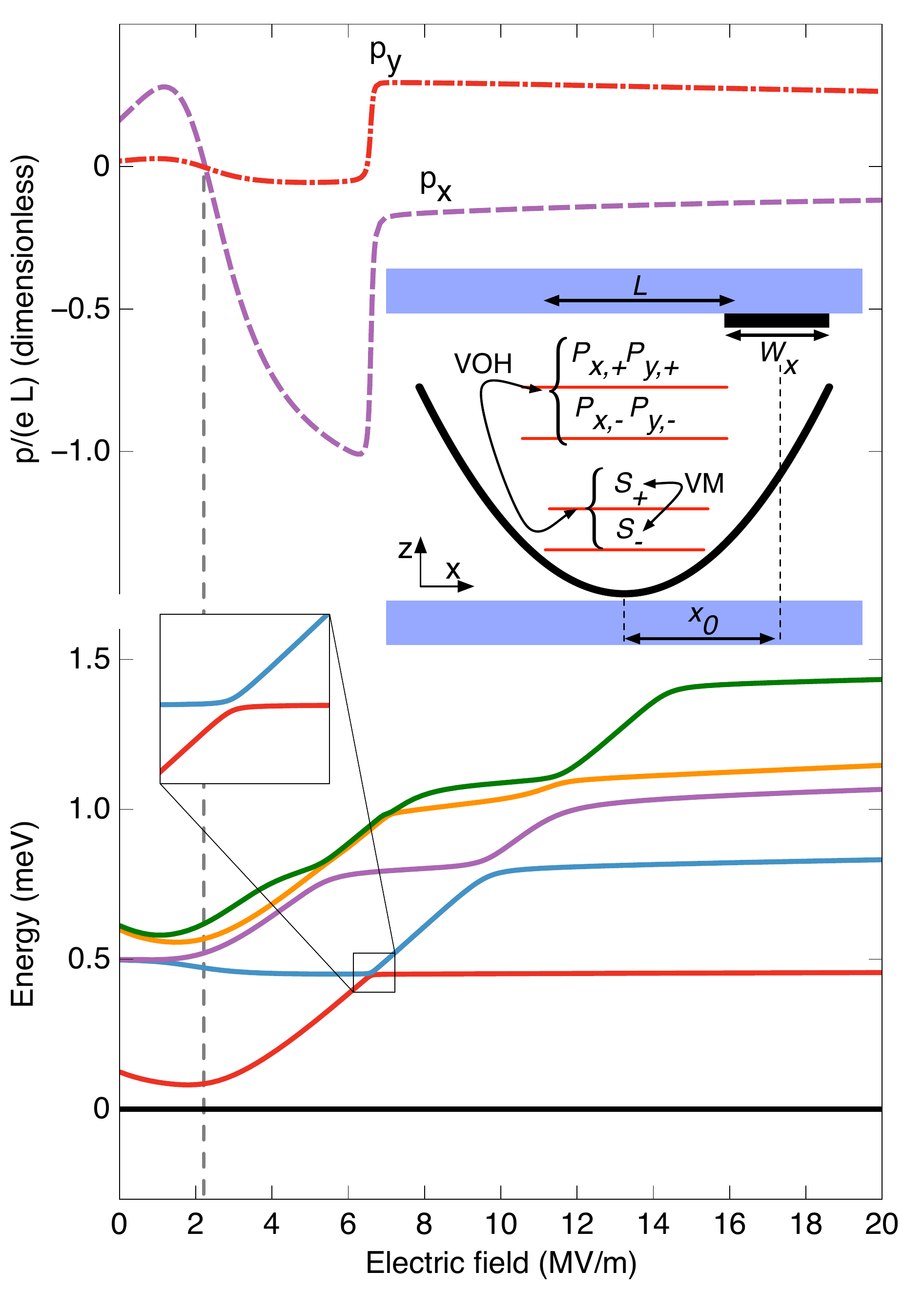}
\vspace{-8pt}
\caption{\label{fig1}(Color online.) 
Calculation of low-lying energy eigenstates and electric dipole moments for a 3D quantum dot in a quantum well with interface disorder, demonstrating that valley-orbit mixing induces a
substantial dipole moment.
The simulation geometry (inset) has a quantum well thickness of $10$~nm and a barrier height of $150$~meV.
The parabolic quantum dot is circular, with a diameter of $L=28.3$~nm. 
Disorder is introduced as a rectangularly shaped ``bump" (black region) in the quantum well barrier, with an $x$-width of $W_x = 2 L$, a $y$-width of $W_y = 4 L$, a height of a single atom, and a center position at $(x_0,y_0) = (-0.7 L,-0.7 L)$. 
The top two curves show the components 
of the dipole moment $\mathbf p$,
defined in Eq.~(\ref{eq:dipole}), 
along the $\hat x$ (dashed) and $\hat y$ (dash-dotted) directions, as a function
of electric field applied along $\hat{z}$. 
The lower set of solid curves show the lowest six energy levels in the quantum dot,
measured relative to the ground state.
At low fields, these form two manifolds:
a lower, $S$-like doublet and an upper $P$-like quadruplet. 
Disorder introduces two distinct effects: valley mixing (VM) between states in the same valley doublet, and valley-orbit hybridization (VOH) between states in different valley doublets.
The dipole moment is typically substantial, but it
is suppressed near the VM-induced anticrossing indicated by the vertical dashed line.
A second anticrossing occurs at a higher field, where
the first excited state changes from valley-like to orbital-like, and is accompanied by a large change in the dipole moment.
}
\vspace{-8pt}
\end{figure}

We now introduce disorder through the perturbation potential $D(\mathbf{r})$.
For the ``bump" geometry shown in Fig.~\ref{fig1}, $D(\mathbf{r})$ is zero everywhere except for the small black region, where it has the same height as the barrier potential.
We write the full Schr\"odinger equation, including the disorder potential, as
\begin{equation}\label{disorder_ham}
\left[ H^0_{xy} + H^0_z + D(\mathbf r) \right]  \phi_l(\mathbf r) = E_l  \phi_l(\mathbf r).
\end{equation}
We solve this equation by expanding in terms of the unpertubed basis set:
\begin{equation} \label{eq:expand}
 \phi_l (\mathbf r)= \sum_{j,k = 1}^{\infty} \alpha^{l}_{jk}  F_j(x,y) \psi_k(z) ,
\end{equation}
where $\alpha^{l}_{jk}$ 
are the expansion coefficients.  
The problem then reduces to a matrix eigenvalue problem for the coefficients $\alpha^{l}_{jk}$ and the energies $ E_l $.

The expansion described in Eqs.~(\ref{disorder_ham}) and (\ref{eq:expand}) must be truncated to find a numerical solution.  
However, the method is guaranteed to succeed if a sufficiently large basis set is used.
Since matrix elements of the disorder potential $D(\mathbf r)$ can be large, we may need many basis functions to obtain quantitative accuracy. In the problems studied below, 
accurate, converged solutions can be obtained reasonably quickly
by using tens of the $\psi_k(z)$ basis functions and tens of the $F_j(x,y)$ basis functions, leading to a
dense effective Hamiltonian matrix with dimensions $N\times N$, where  $N\sim$100-500.

\emph{Application to quantum dot systems.}---We now apply this disorder-expansion technique to calculate the eigenstates of a single quantum dot, and the tunneling coefficients for a double quantum dot, in the presence of a disordered interface. Many previous analytic treatments of disordered interfaces considered only the effects of valley mixing (VM) between the two low-lying valley states. Here, the disorder-expansion method allows us to treat both VM and valley-orbit hybridization (VOH), which describes the mixing of orbital and valley degrees of freedom, and is observed only when the basis includes more than one orbital degree of freedom \cite{Friesen:2010p115324}.
When VOH is significant, the electric dipole moment between the lowest two states,
\begin{equation} \label{eq:dipole}
 \mathbf p = e\int d^3 \mathbf r \left( \left| \phi_1 \left(\mathbf r \right)\right|^2 - \left| \phi_0 \left(\mathbf r \right)\right|^2 \right) \mathbf r ,
\end{equation}
where $-e$ is the electron charge, can also be significant.
In constrast, Eq.~(\ref{eq:dipole}) yields $\mathbf{p}=0$ when $\phi_0$ and $\phi_1$ represent pure pure orbital states,
and is on the scale of the atomic lattice spacing when $\phi_0$ and $\phi_1$ are pure pure valley states.
For qubit applications, a finite dipole moment makes the system susceptible to charge noise \cite{Hayashi:2003p226804,Hu:2006p100501,Gamble:2012p035302}.

We consider the specific quantum dot geometry shown in Fig.~\ref{fig1}.
For simplicity, we choose a 2D parabolic confinement potential for the quantum dot with an energy level spacing of $\hbar \omega = 0.5$~meV, corresponding to a characteristic dot size of $L = \sqrt{\hbar/(m_t \omega)}=28.3$~nm.
We choose the lateral dimensions of the bump perturbation to be of order of $L$, as consistent with recent stuctural characterization of Si/SiGe heterostructures~\cite{Evans:2012p5217}; specifically, we use $W_x=2L$ and $W_y=4L$.
The height of the bump is taken to be a single atom.
The quantum well width is $10$~nm, with a barrier height of $150$~meV.
We compute the $z$-basis functions within the 1D tight-binding method described in Ref.~\cite{Boykin:2004p165325}.
The full 3D calculations are carried out using the disorder-expansion framework described above, with a basis of size $5(x)\times 5(y)\times 30(z)$.
The results are of good accuracy, as described in Appendix~\ref{convergence}.

The results of our single-dot calculations are shown in Fig.~\ref{fig1}  as a function of the perpendicular electric field.
The curves at the top of the plot $p_x$ and $p_y$, the $\hat x$ and $\hat y$ components of the electric dipole moment.
For the device specifications used here,
the dipole moment is typically comparable to $e L$.
Its non-monotonic dependence on perpendicular electric field
 can be understood by examining the energies of the lowest six energy eigenstates of the quantum dot, shown in the lower portion of Fig.~\ref{fig1}. 
At low fields, the lowest set of six energy eigenstates splits into two orbital manifolds.  
Each of these manifolds is further split by a small valley splitting of order 0.1~meV.
As the field is increased, the lowest two states undergo successive transitions, corresponding to level anticrossings: a VM transition at about $2\times 10^6$~V/m, and a VOH transition at about $7\times 10^6$~V/m.
The VM anticrossing is caused by a competition beween two different confinement potentials:  the quantum dot and the effective confinement caused by disorder.
The dipole moments 
are strongly suppressed at the VM
anticrossing, as shown in Fig.~\ref{fig1},
although the magnitude of $\mathbf p$ is never zero.
We note that in the limit of large orbital energy spacing $ \hbar \omega$, the dipole moment 
scales approximately as $1/(\hbar \omega)$, as consistent with lowest order perturbation theory.  
Therefore, two methods are available to help suppress the unwanted dipole moment:
using smaller dots and working at fields corresponding to the VM anticrossing.

\begin{figure}[tb]
\includegraphics[width=8cm]{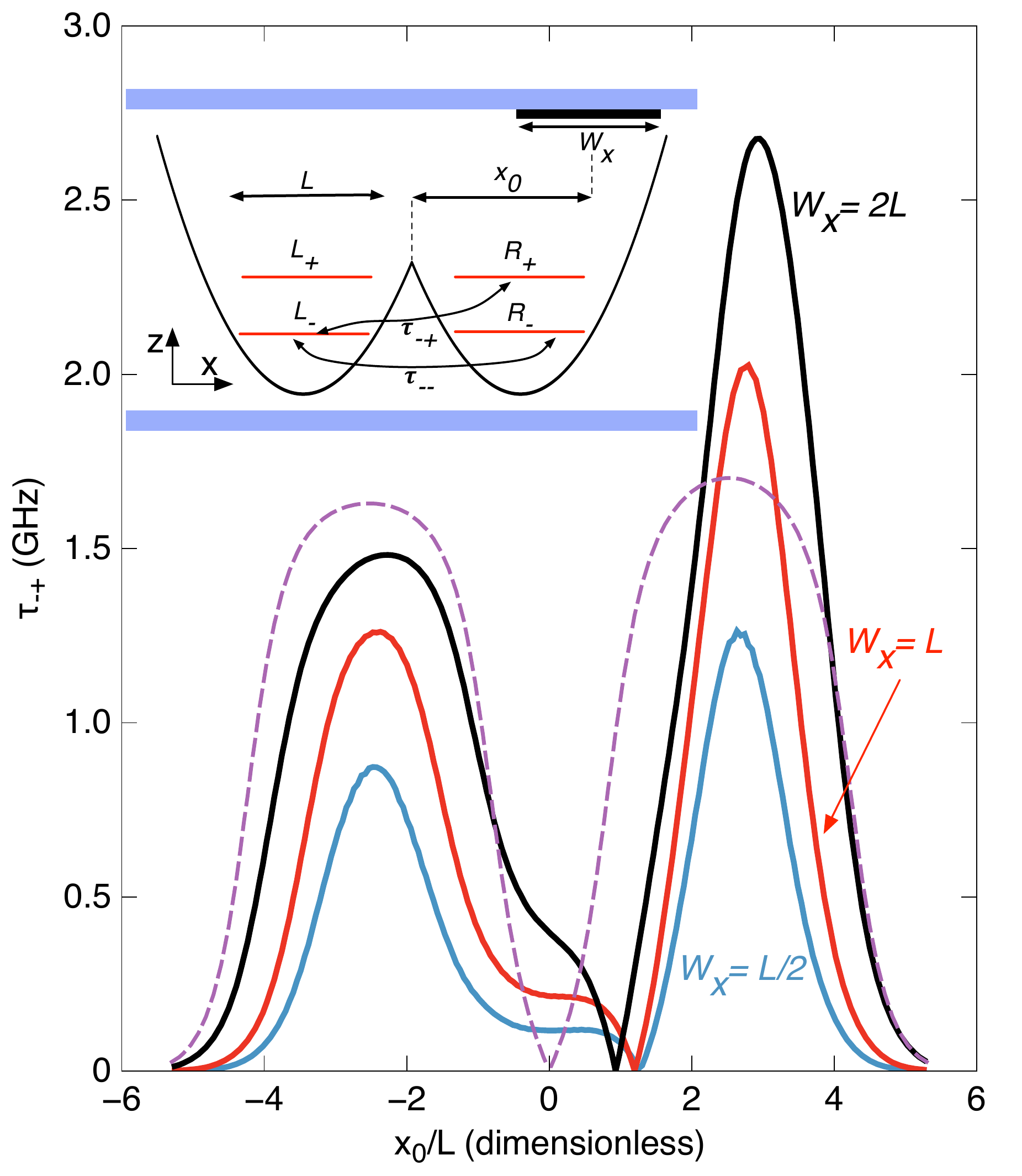}
\vspace{-8pt}
\caption{\label{fig2}(Color online.) 
The intervalley tunnel rate $\tau_{-+} = \left< L_- \right| H \left| R_+ \right>$ between two sides of a double quantum dot in the presence of a ``bump" at the quantum well interface, as a function of the bump position $x_0$.
Here, the height of the bump is one atom, $L_-$ refers to the lowest left-localized state, and $R_+$ refers to the first excited right-localized state.
A schematic of the calculation geometry is shown in the inset.  
In the absence of disorder, the $\pm$ indices refer to unperturbed valley states. 
Results are shown for bumps with widths $W_x=L/2$, $L$, and $2L$ in the $x$-direction, and infinite widths in the $y$-direction.
The solid curves are computed using the disorder-expansion framework described in the main text, with a basis of size $50(x)\times 1(y)\times 10(z)$, while the dashed line is an alternative result for $W=2 L$, with a basis of size $1(x)\times 1(y)\times 2(z)$, which does not admit VOH effects.
For the calculations shown here, the dimensions of the individual dots are the same as Fig.~\ref{fig1}, and we assume an electric field of $2\times 10^5$~V/m applied perpendicular to the quantum well.
The intervalley tunneling rate is substantial over a wide range of bump positions and widths. }
\vspace{-8pt}
\end{figure}

We now study the impact of VOH on interdot tunneling. As noted in Refs.~\cite{Culcer:2010p205315,Friesen:2010p115324}, 
structural disorder induces VM,
so the $z$-component of the wavefunction is no longer well described by its unperturbed eigenstate.
Since disorder varies spatially, the $z$-composition of the wavefunction will be different from dot to dot.
This leads to intervalley tunneling, meaning that an electron can change valley indices when tunneling between two dots \cite{Shiau:2007p195345}.
When quantum information is stored in the valley indices, intervalley tunneling constitutes a loss of information.
Here, we go beyond previous work \cite{Culcer:2010p205315} by 
considering tunneling between the two sides of a double quantum dot,
and by including VOH effects using the disorder-expansion technique described above. 

We again adopt a simple model for interface disorder:  a rectangular bump at the quantum well interface, as shown in the inset of Fig.~\ref{fig2}. 
In this calculation, we take the bump width to be infinite in the $y$ direction, but variable in the $x$ direction.
As anticipated 
in Ref.~\cite{Culcer:2010p205315}, 
the intra- and inter-valley tunnel rates are comparable when
the bump width, analogous to the disorder correlation length, is comparable to the lateral widths of the quantum dot.  We consider two quantum dots in a biquadratic potential.
The individual dot potentials have circular symmetry, with a diameter of $L=28.3$~nm, and an interdot separation of $d = 144$~nm. We define the states $L_{\pm}$ and $R_{\pm}$ to be the left and right-localized states, obtained from the lowest two eigenstates of the left and right individual confinement potentials. All four states $L_{\pm}$ and $R_{\pm}$ are computed within the disorder-expansion framework 
using a basis set of size $50(x)\times 1(y)\times 10(z)$.
(See Appendix~\ref{convergence} for convergence details.%
)

To calculate the tunneling, we compute the matrix element of the total, double dot Hamiltonian between left- and right-localized states. 
Technical details for efficiently computing this matrix element can be found in Appendix~\ref{matrix_elements}.
We run our calculation at a low applied electric field, $F = 2 \times 10^5$~V/m, so that in the absence of disorder, the lowest two states in each dot form a valley doublet. 
This is the regime where the valley index is most likely a good quantum number for quantum computing.
With no disorder, the $x$ and $z$-directions are separable, so the intervalley tunneling term is zero:
\begin{equation}\label{eq:intervalley}
\tau_{-+} = \left< L_- \right| H \left| R_+ \right> = 0~.
\end{equation}
However, 
the introduction of an atomic bump leads to significant intervalley tunneling, as shown in
Fig.~\ref{fig2}. In the calculation, the dot geometry is chosen such that the interdot, intravalley tunnel rate in the absence of disorder is 2~GHz, which is a typical value observed in experiments \cite{Shi:2012preprint}. Over a wide range of bump positions, we confirm that the intervalley tunnel rate is comparable to the intravalley tunnel rate. 
We find that the intervalley tunnel rate is largest when the bump in the interface is centered over one of the quantum dots.
The dashed line in Fig.~\ref{fig2} corresponds to only using one $x$-basis function,  one $y$-basis function, and two $z$-basis functions, corresponding to the simple model considered in previous studies \cite{Culcer:2010p205315}, where the VOH coupling is effectively turned ``off." Although the approximate solution is qualitatively similar to the accurate solution, it is not quantitatively accurate. 

Our theoretical results for intervalley tunneling are in reasonable agreement with recent experiments in a double quantum dot, where tunnel rates were measured between a $(2,1)$ electron occupation and two different $(1,2)$ occupations, corresponding to the ground and lowest excited states~\cite{Shi:2012preprint}.
The small energy splitting between the $(1,2)$ configurations ($\sim 45$~$\mu$eV) is indicative a large valley component in the excited state.  
(Orbital excitations are typically larger, in the range 0.1-1~meV~\cite{Thalakulam:2011p045307}.)
The fact that comparable tunnel rates were observed for both $(1,2)$ states (2.7~GHz vs.\ 3.5~GHz) indicates a strong intervalley matrix element.

Finally, it is interesting to compare the numerical complexity of our scheme to that of a tight binding method.
For the double dot considered here, we achieve good accuracy with a basis of size $N=500$.  
The computational bottleneck in this procedure is diagonalizing the resulting $N\times N$ dense matrix, which takes a few seconds on a personal computer.
In contrast, the number of atoms involved in a 3D tight binding calculation (excluding atoms outside the quantum well) corresponds to including several hundred million atomic sites,
which
requires run-times of many hours on modern supercomputers \cite{Ahmed:2009preprint}.

\emph{Discussion.}---
We have introduced a new disorder-expansion effective mass technique for studying disordered silicon systems. This framework provides results consistent with computationally 
intensive 
tight-binding calculations \cite{nielsen:2013p114304},
while retaining the calculational simplicity and intuitive appeal
of the effective mass approach.
This approach reveals additional valley-orbit hybridization effects, which are responsible for a non-vanishing dipole moment between valley states, as well as intervalley tunneling.

Both 
valley mixing and valley-orbit hybridization
are problematic for storing quantum information in valley states, since in the presence of disorder they no longer afford protection against charge noise, and do not have consistent quantum numbers between dots. We find that the dipole moment can be mitigated by operating the device at a specific applied electric field, and also by making the dot smaller.

The authors acknowledge useful conversations with A.~L.~Saraiva and Belita Koiller. This work was supported in part by ARO and LPS (W911NF-12-0607), and by the United States Department of Defense.
The views and conclusions contained in this document are those of the authors and should not be interpreted as representing the official policies, either expressly or implied, of the US Government.  

\appendix
\appendix

\section{Convergence of the disorder expansion}\label{convergence}
\begin{figure}[tb]
\includegraphics[width= 0.9 \linewidth]{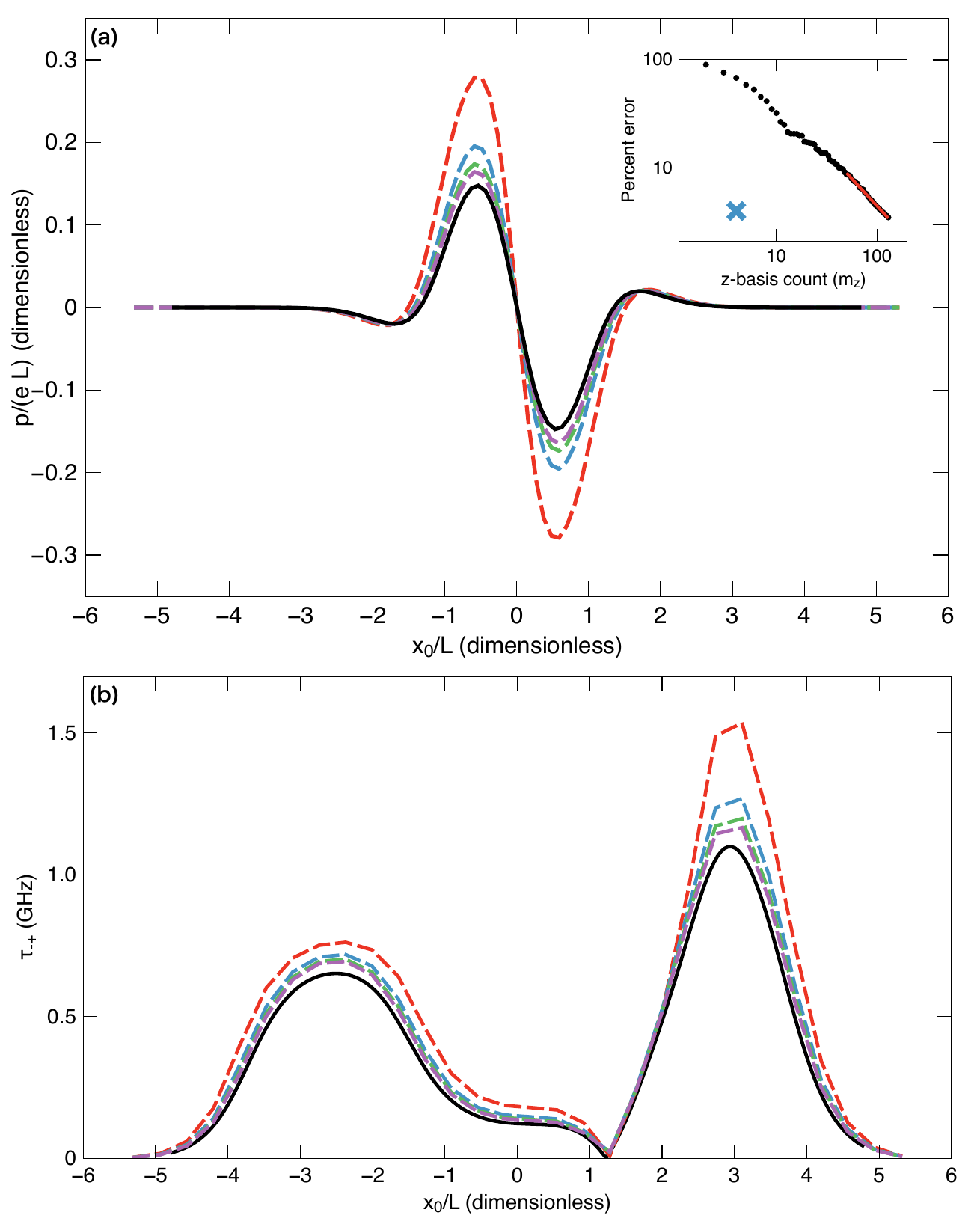}
\vspace{-8pt}
\caption{\label{fig4}
(a): Comparison of disorder-expansion calculations for dipole moment with 2D tight-binding calculations. A bump one atom high in $\hat z$ with a width in $\hat x$ of 300 atoms is centered at the lateral position $x_0$. As in the main text, an electric field is applied along the $\hat z$ direction with strength $2\times 10^5$~V/m, the dot has width $L = 28.3$~nm, and the quantum well is $10$~nm thick with barrier height $150$~meV. The black, solid curve is the tight-binding result, while the colored, dashed lines correspond to different numbers of $z$-basis functions: red is 2, blue is 10, green is 20, and purple is 40. In all cases, 5 $x$-basis functions were used. 
Assuming the tight binding results reflect exact solutions,
the inset shows the percent error in the left peak of the main plot as a function of $m_z$, the number of $z$-basis functions used. The curve fit for large $m_z$ indicates that the percent error scales like $m_z^{-1.005\pm0.007}$ for large $m_z$. The blue cross ($\times$) indicates the percent error obtained using eight $z$-functions from an augmented basis set described in Appendix~A.  (b): Comparison of disorder-expansion calculations for intervalley tunneling with 2D tight-binding calculations. The disorder and system parameters are identical to panel (a), except that two dots are used, and are separated by a distance $d = 150.3$~nm. As in panel (a), the black, solid line indicates tight-binding results, while the colored, dashed lines correspond to successively more $z$-basis functions. Even though tunneling is very sensitive to small tails of the wavefunctions along $\hat x$, the 35 $x$-basis functions used here are sufficient to ensure stability such that the number of $z$-basis functions used limits the accuracy.} 
\vspace{-8pt}
\end{figure}

In this appendix, we examine the convergence properties of the disorder-expansion method introduced in the main text. 
To do this, in Fig.~\ref{fig4} we compare the performance of our disorder-expansion method in a 2D system to results obtained using the tight-binding method of Refs.~\cite{Saraiva:2010p245314,shi:2011p233108}.
Specifically, Fig.~\ref{fig4}~(a) shows the $x$-component of the electric dipole moment $\mathbf p$, defined in Eq.~(\ref{eq:dipole}).
To be able to compare the disorder-expansion results to tight-binding in a reasonable time, we used a 2D system, so we altered the disorder from the single-atom square
described in the main text. 
Analogous to the simple disorder in 3D used in the main text, here we consider a single-atom bump in $\hat z$ with $x$-width $W_x = 70.2$~nm, corresponding to 300 atoms.
This 2D tight-binding problem can be computed in about 10 minutes on a personal computer and compared to the results of the disorder expansion.
As in the main text, an electric field is applied along the $\hat z$ direction with strength $2\times 10^5$~V/m, the dot has width $L = 28.3$~nm, and the quantum well is $10$~nm thick with barrier height $150$~meV.

In Fig.~\ref{fig4}~(a), the black curve is the tight-binding result, while the color lines indicate different numbers of $z$-basis functions: red is 2, blue is 10, green is 20, and purple is 40. In all cases, 5 $x$-basis functions were used.
As more $z$-basis functions are used, the disorder-expansion results become more accurate
(\emph{i.e.}, they approach the tight binding results).
In order to quantify this further, we plot in the inset the percent error in the left peak as a function of $m_z$, the number of $z$-basis functions used. For large $m_z$, we observe that the error falls off like  $m_z^{-1.005\pm0.007}$.

Fig.~\ref{fig4}~(b) shows the intervalley tunneling computed using both disorder-expansion and tight-binding techniques. As in panel (a), the black line corresponds to the 2D tight-binding calculation, while the colored lines corresponding to disorder-expansion calculations with different numbers of $z$-basis functions used. The system parameters used are identical to the dipole calculation, except that the disorder-expansion calculations use 35 $x$-basis functions in all cases, and there are two dots, separated by a distance $d = 150.3$~nm.

As is clear from Fig.~\ref{fig4}, the disorder-expansion technique does converge to the tight-binding results as expected, with the essential physics captured at a modest number of basis functions. However, this convergence can be slow, particulaly with respect to the $\hat z$ basis functions. This is because the perturbations we consider involve large energy scales, which effectively shift the positions of the $\hat z$ energy eigenstates. We are effectively trying to reconstruct this shift in position by including many unperturbed basis states.

We can achieve higher accuracy with fewer basis functions by tailoring our initial choice of basis to the particular type of disorder we include in our system. For example, the types of disorder that we considered in this paper all were of the form of single-atom bumps. This suggests that a better $z-$basis would be to include eigenstates not only of the unperturbed quantum well, but also a quantum well that is narrower by the bump height. Effectively, this means that we are supposing that the true $z-$solution will be a sum of eigenstates of the bare well and eigenstates of the well where the bump covers the entire system. By using this tailored basis, we can achieve very high accuracy with many fewer $z-$basis functions. In the inset of  \ref{fig4}~(a), the blue cross ($\times$) indicates the percent error obtained by using this augmented basis with only eight $z-$functions, demonstrating significantly better accuracy than the ``brute-force" approach with significantly more basis functions. Therefore, a promising direction for future study would be to develop physics-informed, tailored basis sets that can help speed convergence for more general forms of disorder.


\section{Efficient computation of matrix elements}\label{matrix_elements}
The initial, unperturbed basis used for the disorder-expansion calculation is separable in at least $\hat z$ and $\hat x - \hat y$ (and sometimes in $\hat x$ and $\hat y$ individually as well). 
The disorder perturbation mixes the eigenstates of the unperturbed problem, making them no longer separable.
One can then use these 3D states to compute matrix elements of desired operators directly. 
However, we find that in practice the direct computation of these 3D matrix elements is computationally intensive, and often takes longer than the disorder-expansion calculation itself.

To bypass this bottleneck, we exploit the separability of our initial basis states in order to speed up calculation of matrix elements.
While this procedure is not strictly necessary, it enables us to speed up our calculations greatly.
We begin with the calculation of the dipole matrix element, Eq.~(\ref{eq:dipole}) in the main text.
For simplicity of presentation, we show here the computation of only $p_x$; the calculation of $p_y$ follows similarly.
Recalling the definition of the expansion in Eq.~(\ref{eq:expand}), we write
\begin{align}
p_x &= \sum_{j,k,l,m}\int d^3 r x  \Big(\alpha_{jk}^1\alpha_{lm}^1F_jF_l\psi_k\psi_m  -\alpha_{jk}^0\alpha_{lm}^0F_jF_l\psi_k\psi_m \Big) \nonumber \\
& = \sum_{i,j,k} \left(\alpha_{jk}^1\alpha_{lk}^1-\alpha_{jk}^0\alpha_{lk}^0 \right) \int dx dy \cdot x F_jF_l .
\end{align}
Here, we suppress the  arguments of the $F$ and $\psi$ functions for notational simplicity. From this, we see that to compute $p_x$, we can precompute the matrix
\begin{equation}
M^{p_x}_{j,l} = \int dx dy \cdot x F_jF_l ,
\end{equation}
which has dimension equal to the number of $x-y$ basis elements used. Then, computing $p_x$ reduces to a simple sum:
\begin{equation}
p_x = \sum_{j,k,l,m} \left(\alpha_{jk}^1\alpha_{lk}^1-\alpha_{jk}^0\alpha_{lk}^0 \right) M^{p_x}_{j,l}.
\end{equation}
Finally, we note that in the case of an initial basis that is separable in both $\hat x$ and $\hat y$, the formula simplifies even further, since $p_x$ is then diagonal in both $\hat y $ and $\hat z$.

Next, we consider the calculation of the intervalley tunneling matrix elements.
We rewrite Eq.~(\ref{eq:intervalley}) from the main text as 
\begin{align}
\tau_{-+} &= \left< L_- \right| \left( V_\text{DD}(x,y) - V_R(x,y) \right) \left| R_+ \right>  \\
&+ \left< L_- \right| \left( \hat T + V_R(x,y) + V_z (z) + D(\mathbf r) \right) \left| R_+ \right>, \nonumber
\end{align}
where $V_\text{DD}$ is the double-dot potential in the $x-y$ plane, $V_R$ is the right-dot potential, $\hat T$ is the kinetic energy operator, and $V_z$ is the unperturbed potential along $\hat z$. 
We have grouped the second term such that it forms the 
Hamiltonian corresponding to the $\left| R_+ \right>$ eigenstate, which lets us write
\begin{equation}
\tau_{-+} = \left< L_- \right| \left( V_\text{DD}(x,y) - V_R(x,y) \right) \left| R_+ \right> + \epsilon_{R_+}\left< L_- \big| R_+ \right> ,
\end{equation}
where $\epsilon_{R_+}$ is the energy eigenvalue for $\left| R_+ \right>$.
By using the same decomposition technique as we did for the dipole moment calculation, we can write
\begin{equation}
\tau_{-+} = \sum_{j,k,l} \alpha_{jk}^{L_-}\alpha_{lk}^{R_+} \left( M^{A}_{j,l} + \epsilon_{R_+} M^{B}_{j,l} \right),
\end{equation}
where the  matrices are defined by
\begin{equation}
M^{A}_{j,l} = \int dx dy \left( V_\text{DD}(x,y) - V_R(x,y) \right) F^L_jF^R_l ,
\end{equation}
and
\begin{equation}
M^{B}_{j,l} = \int dx dy  F^L_jF^R_l.
\end{equation}
Here, the superscript $L$ and $R$ denote basis functions in the left and right dots, respectively.
As before, these matrices can be precomputed to increase computational speed, reducing the computation of $\tau_{-+}$ to a sum.

\bibliography{citations}

\end{document}